\documentclass{CEAS_EuroGNC_2026}

\usepackage[utf8]{inputenc}

\usepackage{graphicx}
\usepackage{amsmath}
\usepackage[version=4]{mhchem}
\usepackage{siunitx}
\usepackage{array}
\usepackage{longtable}
\usepackage{framed}
\usepackage[most]{tcolorbox}
\usepackage{optidef}
\usepackage{longtable}
\usepackage{hyperref}
\usepackage{aliascnt} 
\usepackage{footnotehyper}
\usepackage{booktabs}
\usepackage{placeins}
\usepackage{printlen}
\usepackage{siunitx}
\usepackage{subcaption}
\usepackage{multirow}
\usepackage{bbm}
\newcommand{\unaryminus}{\scalebox{0.75}[1.0]{\( - \)}}

\newtcolorbox{tcbdoublebox}[1][]{%
  enhanced jigsaw,
  sharp corners,
  colback=white,
  borderline={1pt}{-2pt}{black},
  fontupper={\setlength{\parindent}{20pt}},
  #1
}

\title{Mixed-Integer vs. Continuous Model Predictive Control for Binary Thrusters: A Comparative Study}
\papernumber{086}

\ifpdf
\hypersetup{pdfinfo={
Title={Mixed-Integer vs. Continuous Model Predictive Control for Binary Thrusters: A Comparative Study},
Author={Franek Stark, Shubham Vyas, Jakob Middelberg},
Keywords={model predictive control, mixed integer, satellite control, binary thrusters} 
}}
\fi

\usepackage[acronym,automake=false]{glossaries}
\usepackage{shellesc}

\newacronym{orgl}{ORGL}{Orbital Robotics and GNC Lab}
\newacronym{esa}{ESA}{European Space Agency}
\newacronym{asdr}{ASDR}{Active Space Debris Removal}
\newacronym{reacsa}{REACSA}{REcap-ACrobat-SAtsim}
\newacronym{tvlqr}{TVLQR}{Time-Varying Linear Quadratic Regulator}
\newacronym{lqr}{LQR}{Linear Quadratic Regulator}
\newacronym{mpc}{MPC}{Model Predictive Control} 
\newacronym{pwm}{PWM}{pulse-width modulation} 
\newacronym{lp}{LP}{Linear Program} 
\newacronym{nlp}{NLP}{Non-Linear Program} 
\newacronym{qp}{QP}{Quadratic Program} 
\newacronym{mip}{MIP}{Mixed Integer Program} 
\newacronym{milp}{MILP}{Mixed Integer Linear Program} 
\newacronym{mi}{MI}{Mixed Integer} 
\newacronym{miqp}{MIQP}{Mixed Integer Quadratic Program} 
\newacronym{scip}{SCIP}{Good question what does this mean} 
\newacronym{lcp}{LCP}{Linear Complementarity Program} 
\newacronym{lcqp}{LCQP}{Linear Complementarity Quadratic Program} 
\newacronym{lpcc}{LPCC}{Linear Program with Complementarity Constraints} 
\newacronym{mpcc}{MPCC}{Mathematical Program with Complementarity Constraints} 
\newacronym{lcc}{LCC}{Linear Complementarity Constraints} 
\newacronym{rw}{RW}{Reaction Wheel}
\newacronym{mimpc}{MIMPC}{Mixed Integer Model Predictive Control} 
\newacronym{rms}{RMS}{root-mean-square} 
\newacronym{estec}{ESTEC}{European Space Research and Technology Centre} 
\newacronym{p2pamp}{p-p}{peak-to-peak amplitude} 

\newacronym{recap}{RECAP}{Reaction Control Autonomy Platform}
\newacronym{satsim}{SATSIM}{Satellite Simulator}
\newacronym{acrobat}{ACROBAT}{Air Cushion Robotic Platform}
\newacronym{cftoc}{CFTOC}{Constrained finite time optimal control}
\newacronym{ocp}{OCP}{Optimal Control Problem}
\newacronym{pwpf}{PWPF}{pulse-width pulse-frequency}
\newacronym{cmg}{CMG}{Control Moment Gyro}
\newacronym{mpsp}{MPSP}{Model Predictive Static Programming}
\newacronym{ep}{EP}{Electrical Propulsion}
\newacronym{ds}{$\Delta\Sigma$}{Delta-Sigma}

\begin{document}
\glsdisablehyper

\thispagestyle{fancy}

\maketitle
\pagestyle{empty} 

\begin{authorList}{4cm} 
\addAuthor{Franek Stark \orcidlink{0000-0001-7146-2550}}{Robotics Researcher, Robotics Innovation Center, German Research Center for Artificial Intelligence GmbH\rorlink{https://ror.org/01ayc5b57}, Bremen, Germany. \emailAddress{franek.stark@dfki.de}}
\addAuthor{Jakob Middelberg \orcidlink{0009-0000-7126-3981}}{Research Assistant, Robotics Innovation Center, German Research Center for Artificial Intelligence GmbH\rorlink{https://ror.org/01ayc5b57}, Bremen, Germany. \emailAddress{jakob.middelberg@dfki.de}\newline
Master Student, University of Bremen\rorlink{https://ror.org/04ers2y35}, Germany. \emailAddress{middelberg@uni-bremen.de}}
\addAuthor{Shubham Vyas \orcidlink{0000-0001-8773-8435}}{Robotics Researcher, Robotics Innovation Center, German Research Center for Artificial Intelligence GmbH\rorlink{https://ror.org/01ayc5b57}, Bremen, Germany. \emailAddress{shubham.vyas@dfki.de}}
\end{authorList}
\justifying

\begin{abstract}
Binary on/off thrusters are commonly used for spacecraft attitude and position control during proximity operations. However, their discrete nature poses challenges for conventional continuous control methods. The control of these discrete actuators is either explicitly formulated as a mixed-integer optimization problem or handled in a two-layer approach, where a continuous controller's output is converted to binary commands using analog-to-digital modulation techniques such as \gls{ds}-modulation. This paper provides the first systematic comparison between these two paradigms for binary thruster control, contrasting continuous Model Predictive Control (MPC) with Delta-Sigma modulation against direct Mixed-Integer MPC (MIMPC) approaches.
Furthermore, we propose a new variant of MPC for binary actuated systems, which is informed using the state of the Delta-Sigma Modulator. The two variations for the continuous MPC along with the MIMPC are evaluated through extensive simulations using ESA's \acrfull{reacsa} platform.
Results demonstrate that while all approaches perform similarly in high-thrust regimes, MIMPC achieves superior fuel efficiency in low-thrust conditions. Continuous MPC with modulation shows instabilities at higher thrust levels, while binary informed MPC, which incorporates modulator dynamics, improves robustness and reduces the efficiency gap to the MIMPC.
It can be seen from the simulated and real-system experiments that MIMPC offers complete stability and fuel efficiency benefits, particularly for resource-constrained missions, while continuous control methods remain attractive for computationally limited applications.

\end{abstract}


\section*{Nomenclature}

{\renewcommand\arraystretch{1.0}
\noindent\begin{longtable*}{@{}l @{\quad=\quad} l@{}}
$\theta$  & Platform orientation (yaw) in world coordinates\\
$\dot \theta$  & Platform angular velocity in world coordinates\\
$x, y$  & Platform position in world coordinates\\
$\dot x, \dot y$  & Platform velocity in world coordinates\\
$\omega_{RW}$ & \acrfull{rw} rotational speed \\
$\mathrm{s}_{\theta}, \mathrm{c}_{\theta}$ & Sine and cosine of $\theta$\\
$\textbf{x}, \textbf{u}$ & State and input vector of the system\\
$\hat{\textbf{x}}$ & Target state\\
$F_\text{n}$ & Nominal thrust force applied by a single thruster\\
$r$ & Platform radius\\
$m, I_\text{S}$ & Platform overall mass and inertia on the $z$-axis.\\
$I_\text{RW}$ & \acrshort{rw}'s inertia on the $z$-axis.\\
$\textbf{x}_{j|t}, \textbf{u}_{j|t}$ & State and input prediction of time step $j$, predicted at time step $t$\\
$N$ & Prediction Horizon\\
$\Delta \mathrm{t}$ & Discretization time of the system dynamics within the controller \\
$\textbf{U}_t, \textbf{X}_t$ & Set of all predicted states and inputs at time step $t$\\
$\textbf{u}_{\text{bin},t}$ & Set of all predicted binary inputs at time step $t$\\
$\mathbf{Q}, \mathbf{W}$ & Diagonal cost matrices for state and input cost terms.\\
\end{longtable*}}

\section{Introduction}
Satellites and space vehicles are commonly equipped with either cold gas or hypergolic thrusters for attitude and position control during proximity operations \cite{wie_space_1998}. 
These thrusters are usually binary actuated, i.e., they are on/off devices and can only be either fully on or fully off. 
This is due to the fact that the thrusters are ordinarily designed to operate at a specific pressure and flow rate, and operating them at partial thrust can lead to instability and inefficiency. 
Furthermore, on/off thrusters are often simpler than variable-thrust thrusters, leading to higher reliability and lower cost. 
However, most control methods assume a smooth and continuous state and control input, which is not given with on/off thrusters. 
Therefore, many existing control methods outlined in the literature are not directly applicable, making it challenging to achieve precise attitude and position control.
This results in two options for the development of control systems for proximity operations with binary thrusters. The first is to use a two-layer control architecture, where a continuous controller generates the desired control inputs, and a separate module converts these inputs into binary commands for the thrusters. Different methods have been proposed, from using simple thresholding logic \cite{tsujita_feasible_2023}, up to modulators like \gls{pwm} \cite{banerjee_design_2022}, \gls{ds}-modulation  \cite{zappulla_spacecraft_2017, bredenbeck_trajectory_2022} or similar schemes to convert analog/continuous control input to digital/discrete values. The second option is to directly perform control on the binary control inputs using \gls{mi} methods. Until recently, these were considered too computationally expensive for real-time control, but recent advances in \gls{mi} optimization have made it possible to solve these problems in real-time \cite{sopasakis_hybrid_2015, leomanni_all-electric_2015, stark_linear_2023, stark_mixed_2024}. 
The first approach, while computationally advantageous, can lead to suboptimal performance, as the continuous controller may not be able to fully account for the limitations of the binary thrusters, such as dwell-time constraints (ramp-up or cool-down time). 
The second approach, while more complex, can lead to better performance, as it can directly account for the limitations of the binary thrusters. While earlier work has compared the performance of PWM and \gls{ds}-modulation \cite{zappulla_spacecraft_2017}, to the best of the authors' knowledge, there has been no work comparing the performance of the two-layer approach with \gls{mi} methods for controlling systems with binary thrusters. Such a comparison would allow for understanding the trade-offs between the two approaches and help in understanding if the increased effort for \gls{mi} methods is justified by the performance gains.

In this paper, we compare the performance of a two-layer control architecture using a continuous input \gls{mpc} with \gls{ds}-modulation with a \gls{mimpc} approach for controlling a system with on/off thrusters. 
We also introduce a new \gls{mpc} approach which uses a continuous input \gls{mpc}, that is informed about the behavior of the \gls{ds}-modulator in the prediction horizon. 
This allows the \gls{mpc} to account for the limitations of the binary thrusters while still being able to use efficient numerical methods for solving the continuous optimization problem. 
We compare the performance of these three approaches in terms of their ability to reach and maintain a target position, as well as their efficiency in terms of thruster usage. 
The comparison is performed through a simulation study along with experimental validation of the newly proposed binary-informed \gls{mpc} approach at the Orbital Robotics \& GNC Lab (ORGL) of the \gls{esa} \cite{zwick_orgl_2018}. 
Both the simulation and hardware datasets\footnote{Datasets: \url{https://doi.org/10.5281/zenodo.18454916}}, together with the controller implementation\footnote{Controller implementation: \url{https://github.com/dfki-ric-underactuated-lab/mimpc}}, are publicly available as open source resources.

\section{System Description}
\begin{figure}[htbp]
    \centering
    \begin{subfigure}{0.49\linewidth}
    \includegraphics[width=\linewidth]{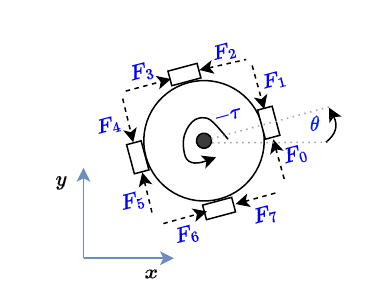}
    \end{subfigure}
    \begin{subfigure}{0.49\linewidth}
    \includegraphics[width=0.8\linewidth, trim={0 0 0 3cm},clip]{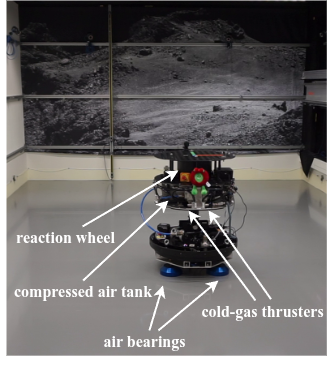}
    \end{subfigure}
    \caption{Sketch (left) of the free-floating platform \gls{reacsa} (right)}
    \label{fig:reacsa}
\end{figure}
The system, which serves as an example for a satellite or small spacecraft, is \gls{esa}'s free-floating platform \gls{reacsa} \cite{suter_reacsa_2023}.
It is a \qty{200}{\kilo\gram} heavy air-bearing platform with a propulsion system containing eight cold-gas thrusters and an additional \gls{rw}.
A sketch of the system, showing the thruster configuration, is shown along with a picture in \autoref{fig:reacsa}.
The cold-gas propulsion enforces a minimum thruster activation time $t_\text{on, min}=\qty{0.1}{\second}$ and limits the maximum firing time of a single thruster to $t_\text{on, max}=\qty{0.3}{\second}$. Further, it enforces a cool-down time of $t_\text{off, min}=\qty{0.2}{\second}$ between consecutive firings of the same thrusters.
\gls{reacsa}'s overall system properties are given in \autoref{tab:reacsa}.
\gls{reacsa} is modeled as a linear time-varying system, the dynamics are given by:
\begin{align}
    \dot{\mathbf{x}} = \underbrace{\begin{bmatrix}
    \multicolumn{3}{c}{\mathbf{0}^{3\times3}}&\multicolumn{3}{c}{\mathbf{I}^{3\times3}}&\mathbf{0}^{3\times1}\\
    \multicolumn{7}{c}{\mathbf{0}^{4\times7}}\\\end{bmatrix}}_{\mathbf{A}}
   \mathbf{x} +
    \underbrace{
             \begin{bmatrix}
      \multicolumn{9}{c}{\mathbf{0}^{3\times9}}\\
        0&\unaryminus \mathrm{s}_{\theta}\frac{F_\text{n}}{m}&\mathrm{s}_{\theta}\frac{F_\text{n}}{m}&\unaryminus \mathrm{c}_{\theta}\frac{F_\text{n}}{m}&\mathrm{c}_{\theta}\frac{F_\text{n}}{m}&
        \mathrm{s}_{\theta}\frac{F_\text{n}}{m}&\unaryminus \mathrm{s}_{\theta}\frac{F_\text{n}}{m}&\mathrm{c}_{\theta}\frac{F_\text{n}}{m}&\unaryminus \mathrm{c}_{\theta}\frac{F_\text{n}}{m}
        \\
        0&\mathrm{c}_{\theta}\frac{F_\text{n}}{m}&\unaryminus \mathrm{c}_{\theta}\frac{F_\text{n}}{m}&\unaryminus \mathrm{s}_{\theta}\frac{F_\text{n}}{m}&\mathrm{s}_{\theta}\frac{F_\text{n}}{m}
        &\unaryminus \mathrm{c}_{\theta}\frac{F_\text{n}}{m}&\mathrm{c}_{\theta}\frac{F_\text{n}}{m}&\mathrm{s}_{\theta}\frac{F_\text{n}}{m}&\unaryminus \mathrm{s}_{\theta}\frac{F_\text{n}}{m}\\
        \frac{\unaryminus 1}{I_\text{S}} & \frac{F_\text{n}r}{I_\text{S}} & \unaryminus  \frac{F_\text{n}r}{I_\text{S}} & \frac{F_\text{n}r}{I_\text{S}} &\unaryminus \frac{F_\text{n}r}{I_\text{S}}&\frac{F_\text{n}r}{I_\text{S}}&\unaryminus \frac{F_\text{n}r}{I_\text{S}}&\frac{F_\text{n}r}{I_\text{S}}&\unaryminus \frac{F_\text{n}r}{I_\text{S}}\\
                 \frac{1}{I_\text{RW}}&  \multicolumn{8}{c}{\mathbf{0}^{1\times8}}\\
    \end{bmatrix}}_{\mathbf{B}(\theta)} \mathbf{u}
\end{align}
with the state vector $\mathbf{x} = \begin{bmatrix}
    x&y& \theta& \dot{x}&\dot{y}&\dot{\theta}&\omega_\text{RW}
    \end{bmatrix}^T$ containing the system's pose, velocity and \gls{rw} velocity.
The input vector is $\textbf{u}_\text{bin} = \begin{bmatrix}
        u_0&u_1&u_2&u_3&u_4&u_5&u_6&u_7&u_8
    \end{bmatrix}^T$, where the first element is the \gls{rw}'s  continuous torque $u_0 \in \mathbb{R}$. The remaining eight elements are the binary thruster activations $u_{1\dots8}\in\left\{0,1\right\}$.
An important fact is that systems with on/off thrusters and a minimum firing time, also known as dwell time constraint, can not always be brought to a full stop. 
Instead, a controller keeps them in a limit-cycle around the target \cite{mendel_performance_1968}. 
\begin{table}[t]
    \centering
    \small
    \caption{\gls{reacsa}'s physical properties}
    \begin{tabular}{@{}llllllllll@{}}
    \toprule
     $m$ & $r$ & $I_s$ & $I_{RW}$ & $\omega_{RW,\text{max}}$ & $F_\text{n}$ & $\tau_{\text{max}}$ &$t_\text{on,min}$&  $t_\text{on, max}$ & $t_\text{off,min}$\\
     \SI{202.81}{\kg} & \SI{0.35}{\meter} & \SI{12.22}{\kg\meter\squared} & \SI{0.047}{\kg\meter\squared} & \num{\pm 250}~RPM & \SI{10.36}{\newton} & \SI{1.44}{\newton\meter} & \SI{100}{\milli\second} & \SI{300}{\milli\second} & \SI{200}{\milli\second}\\
    \bottomrule
    \end{tabular}
    \label{tab:reacsa}
\end{table}

\section{Controllers}
The different \glspl{mpc} compared in this work are all assembled using the following generic \gls{ocp}:
\begin{mini!}
{\substack{\mathbf{u}_0, \hdots, \mathbf{u}_{N-1},\\\mathbf{x}_0, \hdots, \mathbf{x}_N,\\\mathbf{e}_{u,0}, \hdots \mathbf{e}_{u,N-1},\\\mathbf{e}_{x,0}, \hdots \mathbf{e}_{x,N}}}
{\sum^{N-1}_{t=0} \left( \mathbf{Q}^T\mathbf{e}_{x,t} + \mathbf{W}^T\mathbf{e}_{u,t} \right) + \mathbf{Q_\text{f}}^T\mathbf{e}_{x,N}\label{eq:controller:mpc-cost}}
{\label{eq:controller:mpc-form}}
{}
\addConstraint{\mathbf{x}_{t+1}}{=\mathbf{x}_{t} + \Delta \mathrm{t} ~ \mathbf{A} \mathbf{x}_{t}+ \Delta \mathrm{t} ~ \mathbf{B}(\theta_t)~ \mathbf{u}_{t}, \quad && \forall t \in \{0,\dots,N-1\}\label{eq:controller:mpc-form-dynm}}
\addConstraint{-\mathbf{e}_{x,t}}{\leq ~\mathbf{x}_{t} - \hat{\mathbf{x}} \leq + \mathbf{e}_{x,t}, \quad && \forall t \in \{0,\dots,N\}  \label{eq:l1_cons_x}}
\addConstraint{-\mathbf{e}_{u,t}}{\leq ~\mathbf{u}_{t} \leq + \mathbf{e}_{u,t}, \quad && \forall t \in \{0,\dots,N-1\}  \label{eq:l1_cons_u}}
\addConstraint{-\tau_\text{max}}{\leq \textbf{u}_{t,0} \leq \tau_\text{max}, \quad &&\forall t \in \{0,\dots,N-1\}\label{eq:controller:mpc-form-torque-cons}}
\addConstraint{0}{\leq \textbf{u}_{t,1 \dots 8} \leq 1, \quad &&\forall t \in \{0,\dots,N-1\}\label{eq:controller:mpc-form-bin-limit}}
\addConstraint{\mathbf{x}_\text{lb}}{\leq \textbf{x}_{t} \leq \mathbf{x}_\text{ub},\quad && \forall t \in \{0,\dots,N-1\}\label{eq:controller:mpc-form-state-cons}}
\addConstraint{\mathbf{x}_\text{f,lb}}{\leq \textbf{x}_{N} \leq \mathbf{x}_\text{f,ub}\label{eq:controller:mpc-form-state-final-cons}}
\addConstraint{\mathbf{x}_{0}}{= \mathbf{x}_\text{estimated}.\label{eq:controller:mpc-form-curr-state}}
\end{mini!}
where $N$ is the prediction horizon and $t$ the discrete-time step index.
State vector $\textbf{x}_{t}$ and input vector $\textbf{u}_{t}$ refer to the $t^\text{th}$ future prediction.
The diagonal weight matrices $\mathbf{Q}, \mathbf{Q}_f$ and $\mathbf{W}$ weight the absolute input and state error values over the prediction horizon.
The auxiliary vectors $\mathbf{e}_{x,t}$ and $\mathbf{e}_{u,t}$ are constraint by \eqref{eq:l1_cons_x} and \eqref{eq:l1_cons_u} to take the absolute value of the state error (to target state $\mathbf{\hat{x}}$) and inputs respectively.
By doing so, the \gls{ocp} adopts an $\ell_1$-norm objective and can be cast as a \gls{lp}, which is straightforward to solve.
The forward-Euler discretized and linearized system dynamics are enforced by \eqref{eq:controller:mpc-form-dynm}.
The \gls{rw} input torque is constrained by \eqref{eq:controller:mpc-form-torque-cons} and the thruster inputs are limited to stay between no-thrust ($0$) and full-thrust ($1$) by \eqref{eq:controller:mpc-form-bin-limit}.
The flatfloor bounds, the \gls{rw}'s, and the system's velocity are constrained within the state-space by~\eqref{eq:controller:mpc-form-state-cons}. The recursive feasibility of the \gls{ocp} is ensured by constraining the terminal state $\textbf{x}_{N}$ via \eqref{eq:controller:mpc-form-state-final-cons} to be within a control invariant set as derived in \cite{stark_mixed_2024}.
The first state $\mathbf{x}_0$ is constrained to the current state estimate by \eqref{eq:controller:mpc-form-curr-state}.
The three \glspl{mpc} that are derived from \eqref{eq:controller:mpc-form} are listed in \autoref{tab:mpcs} and explained in the following.

\begin{table}[b]
    \small
    \centering
    \caption{Different \glspl{mpc} compared in this work}
    \begin{tabular}{@{}lll@{}}
    \toprule
        Controller &  Description & Solver \\
         \hline
        \emph{MIMPC} & Enforces binary thruster inputs and timings explicitly & SCIP \cite{bolusani_scip_2024} \\
        \emph{continuous MPC} & Continuous MPC with \acrshort{ds}-modulator  & CLP \cite{john_forrest_coin-orclp_2024} \\
        \emph{binary informed MPC} & Continuous MPC + information on \acrshort{ds}-modulator firings in OCP & CLP \cite{john_forrest_coin-orclp_2024} \\
     \bottomrule
    \end{tabular}
    \label{tab:mpcs}
\end{table}

The first controller is the \textit{\gls{mimpc}} which handles the binary and timing constraints explicitly in the \gls{ocp}. 
By doing so, the \gls{ocp} \eqref{eq:controller:mpc-form} becomes a \gls{mip} which is NP-hard to solve. 
Previous analyses have shown that often the optimal solution or at least a sub-optimal solution can be found for $N=20$ within \qty{0.1}{\second}, hence allowing real-time control with \qty{10}{\hertz} \cite{stark_linear_2023, stark_mixed_2024}.
The details on the \gls{mimpc} can be found in \cite{stark_mixed_2024}.

The second controller is the \textit{continuous \gls{mpc}}. Here the \gls{ocp} \eqref{eq:controller:mpc-form} stays as stated. 
The \gls{lp} can be solved very efficiently, enabling a control loop frequency of more than \qty{100}{\hertz}. 
Consequently, the \gls{mpc} outputs continuous thruster activation values and does not respect the thruster timing constraints. 
A~\gls{ds}-modulator converts the continuous inputs into on/off values. Each thruster has its own modulator which switches the thruster state based on the integrated thruster error ${u}_{j,\text{error}}$ for each thruster $j=1\dots8$:
\begin{subequations}
\begin{align}
    \dot{u}_{j,\text{error}} &= K \left(u_{0,j} - u_{j,\text{bin}}\right)\\
    u_{j,\text{bin}} &= \begin{cases}
    1, & {u}_{j,\text{error}} > \epsilon\\
    0, & \mathrm{else}
    \end{cases}
\end{align}
\end{subequations}
where $u_{0,j}$ is the continuous thruster input for the current timestep $t=0$ determined by solving \eqref{eq:controller:mpc-form}. The actual binary value $u_{j,\text{bin}}$ is the modulator's output.
The integrator gain $K$ is set to $1$ and the trigger threshold $\epsilon$ is set to the minimum activation time $t_\text{on, min}$ for simplicity.
Additional thruster timing logic enforces the thruster timing constraints;
as soon as the modulator fires, the thrust is maintained for the minimum time. After the maximum thrust firing time, the modulator will stop firing, and after each firing, the cool-down time is enforced.

To compensate for the fact that there is now an unmodeled sub-system in the control loop, this work introduces the \textit{binary informed MPC} as the third controller.
The full \gls{ds}-modulator cannot be modeled without making \eqref{eq:controller:mpc-form} a \gls{mip} again.
Instead, we only inform the continuous MPC about the current state of the \gls{ds}-modulator.
This is achieved by simulating the behavior of the \gls{ds}-modulators forward along the prediction horizon $N$ at each control cycle using the above logic and the current thruster errors.
This results for each thruster $j$ in the \gls{ds}-modulator-input series $u_{t,j,\text{bin}},~t = \left\{0,\dots,N-1\right\}$.
Based on this series the dynamic constraint \eqref{eq:controller:mpc-form-dynm} is altered to include the planned \gls{ds}-modulator firings:
\begin{equation}
    \mathbf{x}_{t+1}=\mathbf{x}_{t} + \Delta \mathrm{t} ~ \mathbf{A} \mathbf{x}_{t}+ \Delta \mathrm{t} ~ \mathbf{B}(\theta_t)~ \mathbf{u}_{t} + \Delta \mathrm{t} ~ \mathbf{B}(\theta_t)~ \mathbf{u}_{t,\text{bin}}, \quad \forall t \in \{0,\dots,N-1\}\label{eq:controller:mpc-form-dynm-informed}
\end{equation}
where $\mathbf{x}_t$ and $\mathbf{u}_t$ remain the decision variables of the \gls{ocp} and $\mathbf{u}_{t, \mathrm{bin}}$ is not a decision variable but the stacked vector of the forward simulated modulator behavior for all thrusters at the respective predicted timestep $t$:
\begin{equation}
    \mathbf{u}_{t, \mathrm{bin}}=\begin{bmatrix}
0&u_{t,1,\mathrm{bin}}&u_{t,2,\mathrm{bin}}&u_{t,3,\mathrm{bin}}&u_{t,4,\mathrm{bin}}&u_{t,5,\mathrm{bin}}&u_{t,6,\mathrm{bin}}&u_{t,7,\mathrm{bin}}&u_{t,8,\mathrm{bin}}\end{bmatrix}
\end{equation}

Overall, these three controllers represent a spectrum of how tightly binary actuation is coupled to the optimization problem. In the remainder of this work, we compare these three levels of coupling in terms of computational effort and closed-loop performance under identical experimental conditions. In all three, the horizon is set to $N=20,$ and the discretization step to $\Delta \mathrm{t}=\qty{0.1}{\second}$.


\section{Efficiency Analysis}
To assess each controller's theoretical best performance, each controller was evaluated over $\num{5400}$ simulation experiments.
In each experiment, the system is initialized from the common initial state
\begin{equation}
\mathbf{x}_0 = [1.0,\,-0.5,\,\pi,\,0.0,\,0.1,\,0,\,0].
\end{equation}
The weighting matrices in \eqref{eq:controller:mpc-cost} are chosen as
\begin{align}
    \mathbf{Q}&=
    \mathrm{diag}\!\left(1,\,1,\,0.12,\,\eta,\,\eta,\,0.12\eta,\,0\right),\quad
    \mathbf{Q}_f = \xi\,\mathbf{Q},\\
    \mathbf{W}&=
\mathrm{diag}\!\left(0.0001,\,\kappa,\,\kappa,\,\kappa,\,\kappa,\,\kappa,\,\kappa,\,\kappa,\,\kappa\right).
\end{align}
For each experiment, a single realization of \(\eta\), \(\xi\), and \(\kappa\) is sampled and then applied identically to all three controllers. The sampled parameters follow
\begin{equation}
    \eta \sim \mathcal{U}(0,0.5), \quad
    \xi \sim \mathcal{U}(1,21), \quad
    \kappa \sim \mathcal{U}(0,0.6).
\end{equation}
Each controller's target is to reach the origin, i.e., $\hat{\mathbf{x}}=\left[0,0,0,0,0,0,0\right]$.
An experiment is deemed successful when the system remains within a \qty{0.1}{\meter} radius of the origin for a minimum of \qty{40}{\second}.
Hereby, the value of \qty{0.1}{\meter} is a result from \cite{stark_mixed_2024} in which we derived the theoretical limit cycle bounds for this system, multiplied by a relaxation factor of $3$.
The experiment is considered to be a failure if the target has not been reached after \qty{80}{\second}.
Upon successful completion, we report the time taken to reach the target and the average Euclidean position error of the system at the target.
To determine the efficiency, the \textit{average thruster usage} is defined as the mean duty cycle across all thrusters.
This is calculated by summing each thruster’s total activation time over the duration of the experiment and normalizing it by the product of the number of thrusters and the total experiment duration. 
The average thrust usage is determined individually for reaching the target and staying at the target.
The simulated experiments are carried out under real-time conditions using the Drake toolbox \cite{tedrake_drake_2019}. 
In simulation, the controllers run on an x86 desktop system with   an \textit{i9-10900K} CPU with \num{10} cores, \qty{3.7}{\giga\hertz}. 
As noted earlier, the LP can be solved within a few milliseconds, hence the continuous and binary informed \glspl{mpc} are simulated at a control frequency of \qty{100}{\hertz}. In contrast, solving the \gls{mimpc} is NP-hard, so a control frequency of only \qty{10}{\hertz} is targeted, and the solver is terminated after \qty{0.1}{\second} with a possibly suboptimal solution, as described in \cite{stark_mixed_2024}.
\begin{figure}[htbp!]
    \centering
    \captionsetup{justification=justified, margin=0.5cm}
    \includegraphics[width=\linewidth]{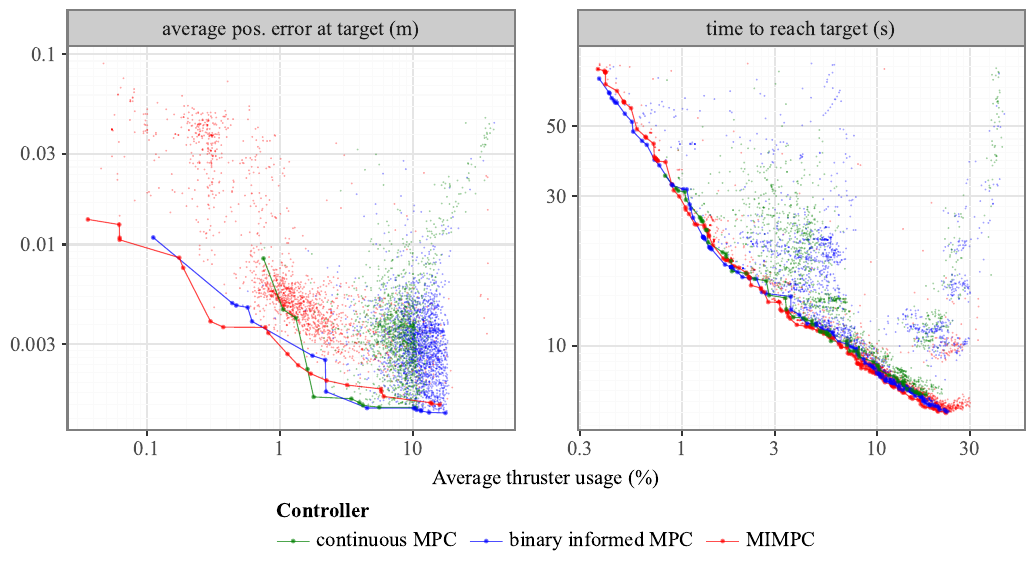}
    \vspace{-1cm}
    \caption{Average position error after reaching and while staying at the target position (left) and time to reach the target position (right) for all compared controllers plotted over the average thrust usage. Single experiments are indicated by small dots, and the Pareto-optimal experiments for all controllers are depicted by solid lines. }
    \label{fig:exp_pareto}
\end{figure}

To assess the controllers' efficiencies, the time to reach the target and the position error when staying at the target are plotted along with the average thruster usages in \autoref{fig:exp_pareto}. The Pareto efficient experiments for each controller are highlighted and connected to a Pareto front.
When comparing the \textit{time to reach the target}, all controllers show similar efficiency as the Pareto fronts overlay.
However, as indicated by the small dots representing non-Pareto-optimal experiments, the continuous \gls{mpc} and the binary informed \gls{mpc} scatter much higher into the inefficient region.
On the low thrust side of the plot, the continuous \gls{mpc} needs an average thruster usage of at least \qty{8}{\percent} to reach the goal, which in this case, takes \qty{34.8}{\second}. 
The \gls{mimpc} and the binary informed \gls{mpc}, in contrast, can use fewer thruster activations of minimum \qty{3.7}{\percent} (\gls{mimpc}) and \qty{3.8}{\percent} (binary informed \gls{mpc}) and take \qty{76.0}{\second} and \qty{70.9}{\second} but still reach the target.
On the high-thrust side, with a thrust usage of \qty{18}{\percent}, the continuous \gls{mpc} achieves its minimum time of \qty{7}{\second} to reach the target.
With higher thrust usage and depending on tuning, the continuous \gls{mpc} still reaches the goal without improving its performance. 
Depending on the tuning, however, it often becomes unstable and in multiple cases even exceeds the flat-floor limits.
In contrast, the \gls{mimpc} and the binary informed \gls{mpc} can further reduce the time to reach the goal with higher thruster usage.
The average thruster activations are \qty{22.6}{\percent} (\gls{mimpc}) and \qty{22.8}{\percent} (binary informed \gls{mpc}) with a minimum time to reach the target of only \qty{6.1}{\second} (\gls{mimpc}) and \qty{6.2}{\second} (binary informed \gls{mpc}).

\begin{figure}
    \centering
    \includegraphics[width=0.8\linewidth]{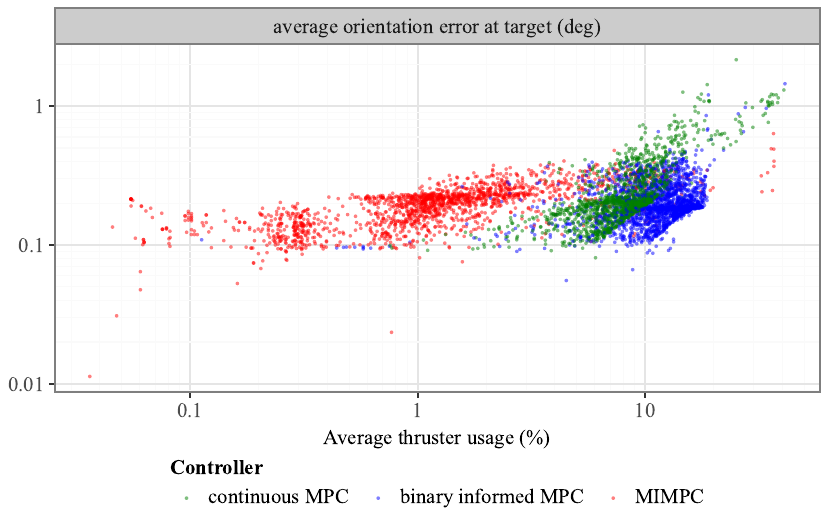}
    \vspace{-0.2cm}
    \caption{Average orientation of all controllers after reaching and while doing station-keeping.}
    \label{fig:comp_soler_orient}
\end{figure}
For the performance at the target (i.e. station-keeping), the difference in efficiencies is more distinct.
The lowest average position error of \qty{1.3}{\milli\meter} is reached by the binary informed \gls{mpc} with an average thruster activation of \qty{17.6}{\percent}. Again, the continuous \gls{mpc} does not reach the goal with this high thruster usage and therefore only reaches a minimum error of \qty{1.4}{\milli\meter} at a thruster usage of \qty{10.5}{\percent}.
Notably, comparing the \gls{mimpc} in this high thrust region of roughly \qty{>2}{\percent}, its Pareto curve lies approximately \qtyrange{0.1}{0.3}{\milli\meter} above the one of continuous and binary informed \gls{mpc}. This indicates a slightly lower efficiency in this region for the \gls{mimpc}.
However, at a thrust usage of roughly \qty{2}{\percent} this switches (for the binary informed \gls{mpc} at a higher thrust usage than the continuous \gls{mpc}, making the latter more efficient in this region).
For thruster usages below \qty{2}{\percent}, the Pareto front of the \gls{mimpc} remains below that of both binary informed and continuous \gls{mpc} except for two outliers in which the Pareto fronts of \gls{mimpc} and binary informed \gls{mpc} intersect.
The minimum thrust usage of the continuous \gls{mpc} at the target is \qty{0.8}{\percent}, with an average position error of \qty{8.4}{\milli\meter}.
Both the \gls{mimpc} and the binary informed \gls{mpc} reach this error, with a thrust of roughly \qty{0.2}{\percent}, making them more efficient.
Ultimately, the binary informed \gls{mimpc} reaches minimum thrust usage of \qty{0.1}{\percent}, with an average position error of \qty{1.1}{\centi\meter}, whereas the \gls{mimpc} achieves the same error with only \qty{0.06}{\percent} of thruster usage.
The lowest thrust usage of \qty{0.04}{\percent} at the target is achieved by the \gls{mimpc}, where it has an average position error of \qty{1.4}{\centi\meter}.
Notably, in the lower thrust region, the points on the Pareto fronts of the respective controllers are relatively sparse.
For the continuous and the binary informed \gls{mpc}, even the Pareto non-optimal experiments become very sparse in this region.  

The average orientation errors at the target for all experiments are depicted in \autoref{fig:comp_soler_orient}.
For an average thruster usage of \qty{<5}{\percent}, all controllers have an average error of less than \qty{0.5}{\degree}.
However, since the firing of a single thruster always affects the orientation,  higher thruster usage leads to an increase in orientation error, reaching up to approximately \qty{1.8}{\degree}. 


\FloatBarrier
\section{Evaluation on the Real System}
\captionsetup{justification=centering,margin=2cm}
\begin{figure}[htbp]
    \centering
    \begin{subfigure}[t]{0.24\linewidth}
        \includegraphics[width=\linewidth]{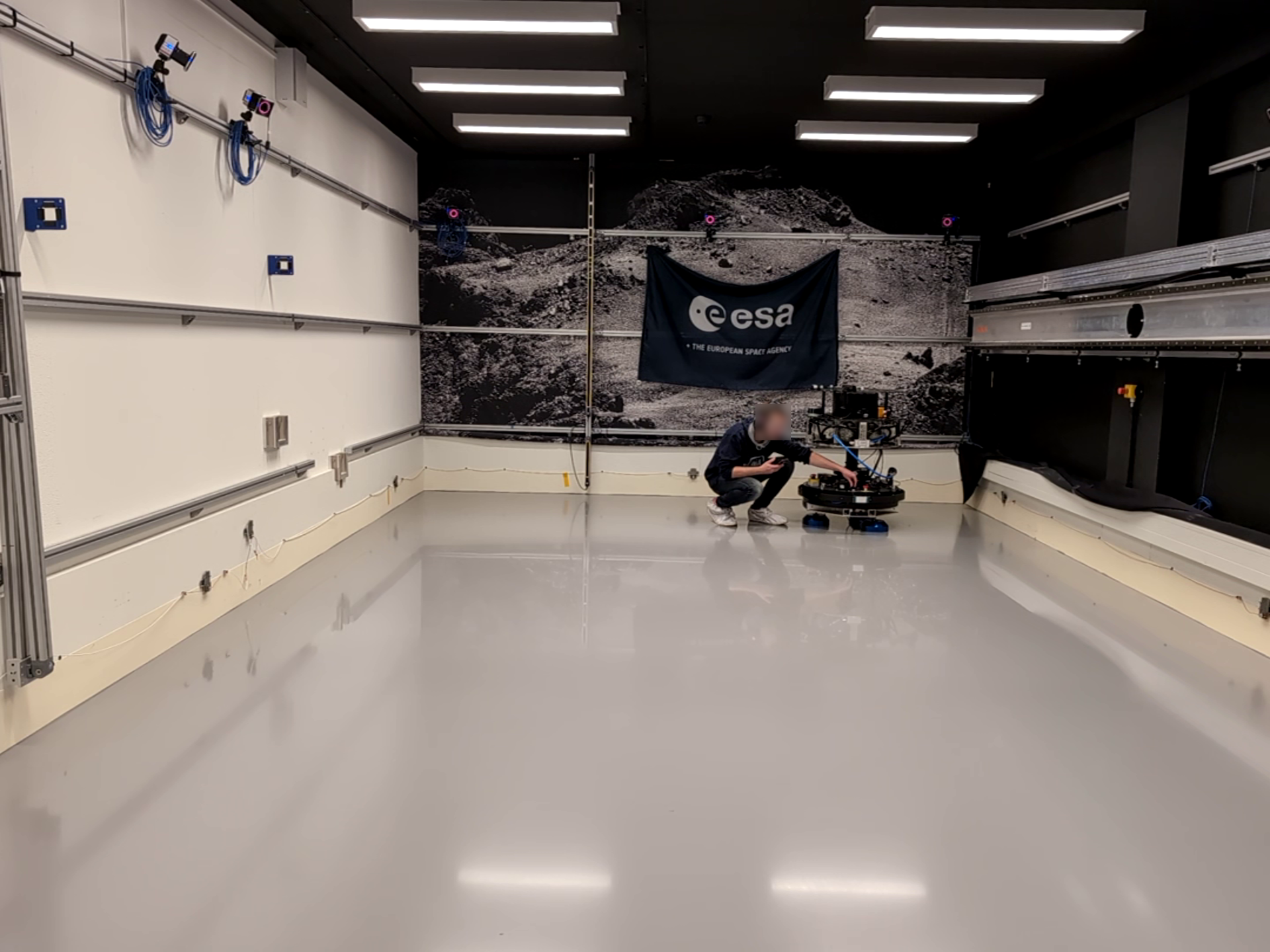}
    \end{subfigure}
    \begin{subfigure}[t]{0.24\linewidth}
        \includegraphics[width=\linewidth]{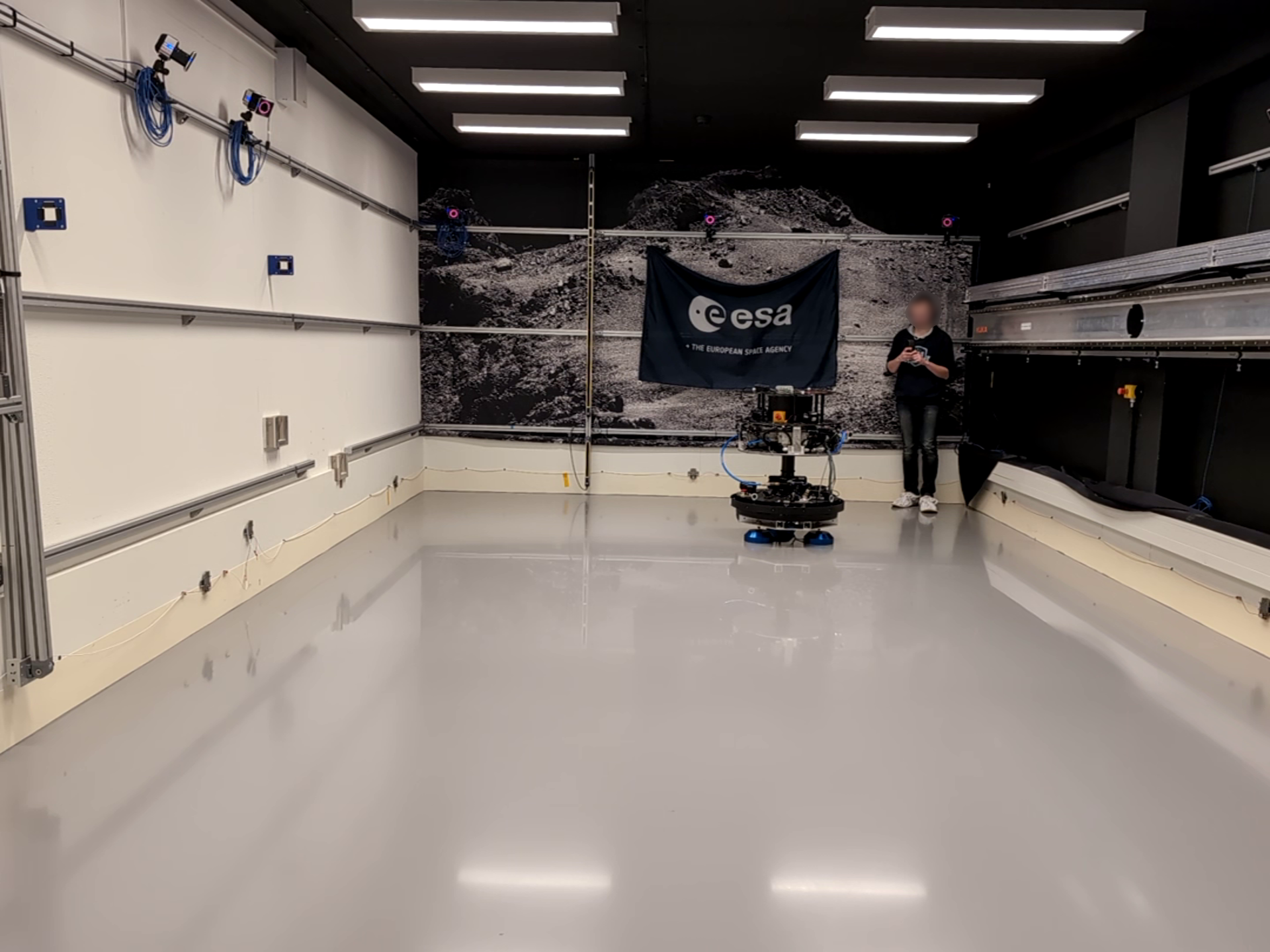}
    \end{subfigure}
    \begin{subfigure}[t]{0.24\linewidth}
        \includegraphics[width=\linewidth]{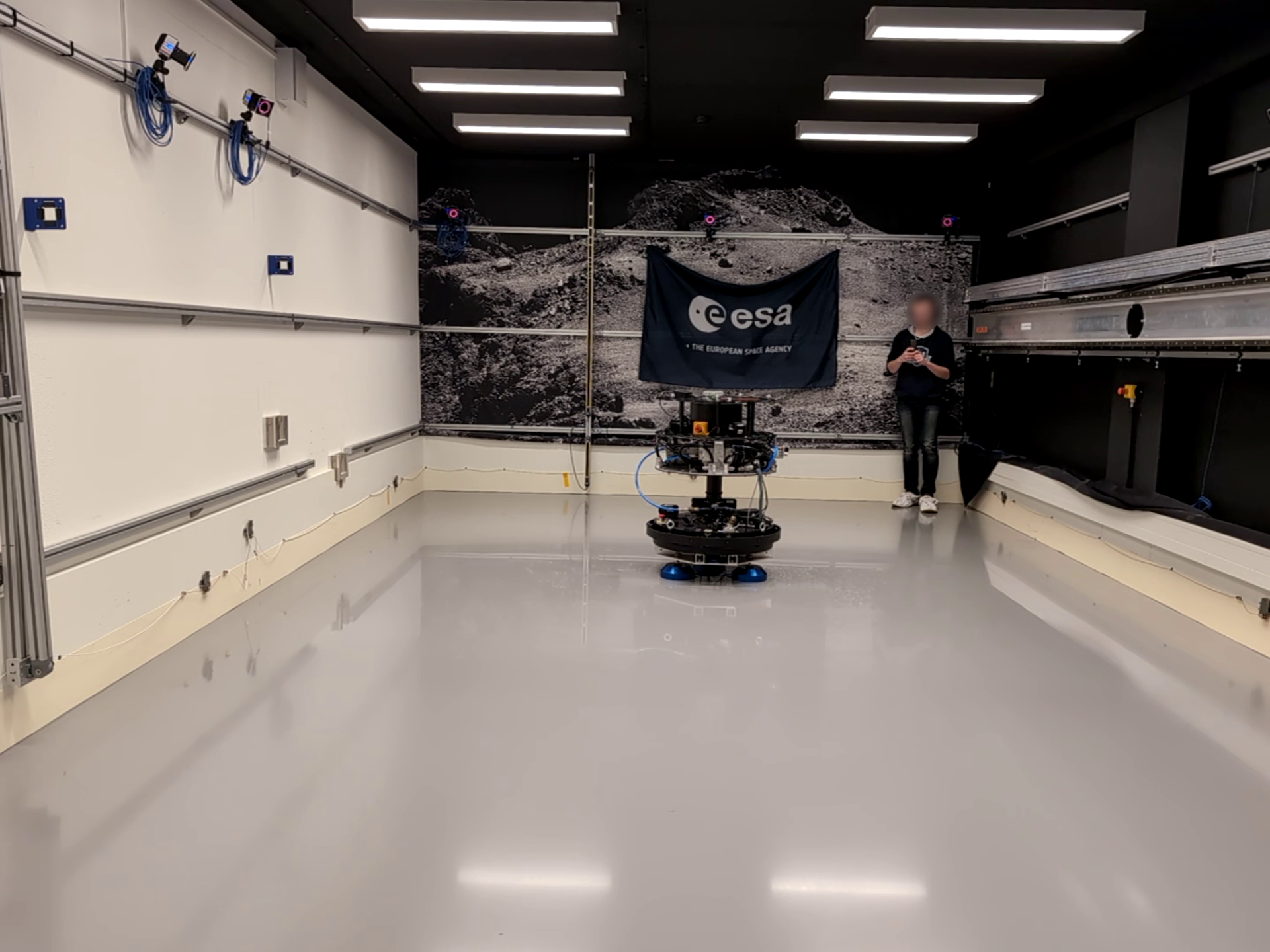}
    \end{subfigure}
    \begin{subfigure}[t]{0.24\linewidth}
        \includegraphics[width=\linewidth]{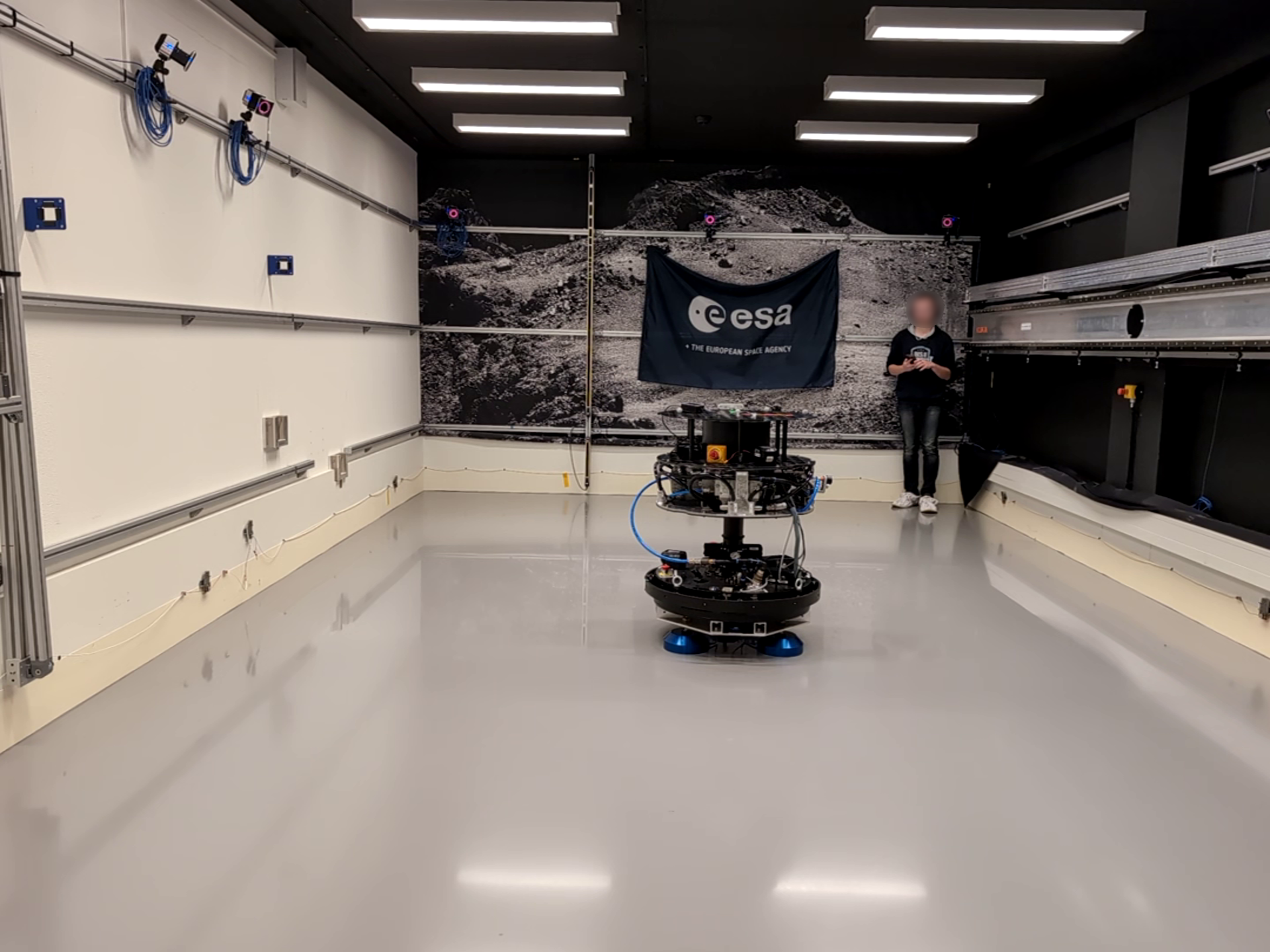}
    \end{subfigure}
    \caption{Carrying out experiments at the ORGL - Images show the transition of the platform from the starting point to the origin in reading direction}
    \label{fig:one_test_procedure}
\end{figure}
\captionsetup{justification=justified,margin=.5cm}
\begin{figure}[htbp]
    \centering
    \begin{subfigure}[t]{0.63\linewidth}
    \includegraphics[width=\linewidth]{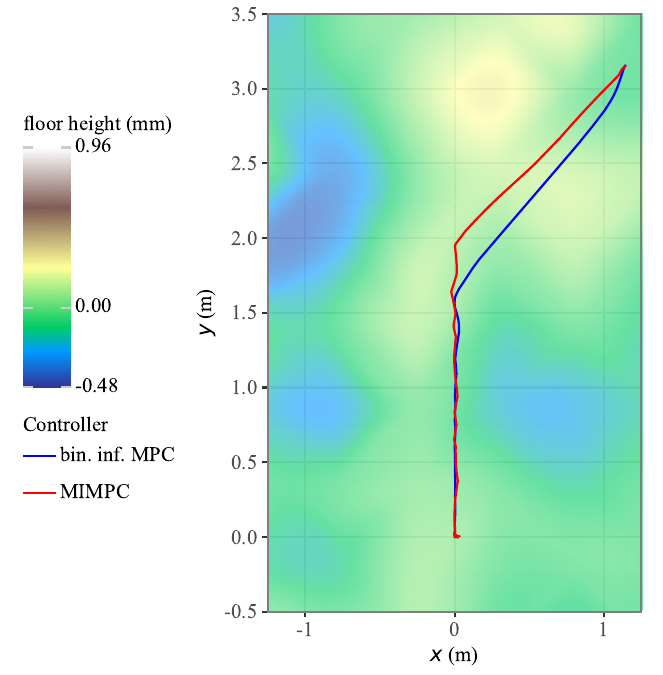}
    \end{subfigure}~
    \begin{subfigure}[t]{0.34\linewidth}
    \includegraphics[width=\linewidth]{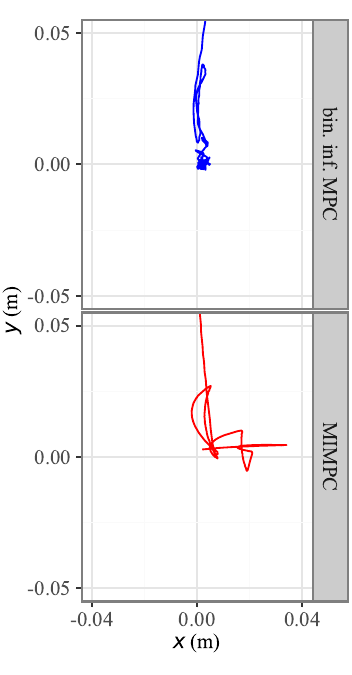}
    \end{subfigure}
    \caption{Resulting trajectory of \gls{reacsa} under control of the \gls{mimpc} and the binary informed \gls{mimpc}.
    The left plot shows the whole trajectory along with the small height variations of the otherwise flat floor, which introduce disturbances.
    The right plots zoom into the limit cycle at the target.
    }
    \label{fig:traj_resl}
\end{figure}
Since the binary informed \gls{mpc} outperforms the purely continuous \gls{mpc} in the previous analysis and the latter was unstable in certain scenarios, only the \gls{mimpc} and the binary informed \gls{mpc} are evaluated and compared on the real system.
In both experiments, the platform is positioned at the starting point $(x=\qty{1.14}{\meter}, y=\qty{3.14}{\meter}, \theta=\qty{90}{\degree})$.
An operator then activates the air bearings manually, followed by activating the respective controller.
The goal is to reach the origin at the floor's center point and de-rotate the platform to $\theta=\qty{0}{\degree}$.
After \gls{reacsa} has reached the origin, the controllers are kept active for another \qty{20}{\second} to study their limit cycle behaviors.
An exemplary experiment is shown in \autoref{fig:one_test_procedure}.
The controllers have been both manually tuned with the same weights.

\begin{table}[htbp]
    \centering
    \begin{tabular}{rlrr}
        \toprule
        && MIMPC &  binary informed MPC  \\
        \midrule
        \multirow{3}{5em}{\textit{Reaching the target}}&Time to reach (s) & 51.268 & 25.209 \\
        &Total thrust usage (s) & 4.90 & 7.51 \\
        &Average thrust usage (\%) & 1.2 & 3.7 \\
        \midrule
        \multirow{3}{5em}{\textit{Limit cycle at the target}}&Average position error (cm) & 1.824  & \textbf{0.853} \\
        &Average orientation error (deg) & \textbf{0.176} & 0.419 \\
        &Average thrust usage (\%) & \textbf{1.4} &  6.0 \\
        \bottomrule
    \end{tabular}
    \vspace{.3cm}
    \caption{Results from experiments on the real system}
    \vspace{-.5cm}
    \label{tab:mpc_results}
\end{table}

The resulting trajectories of \gls{reacsa} under both control laws are shown in \autoref{fig:traj_resl}. \autoref{tab:mpc_results} contains the performance measures respectively. Videos of the experiments are available online\footnote{\url{https://youtu.be/briBzeDHmGs}}. 
Due to the L1-norm cost function, both controllers follow an L-shape towards the goal, as it encourages minimizing the error on all axes independently.
The \gls{mimpc} follows a more rigid L-shape compared to the binary informed \gls{mpc}.
Overall, the binary informed \gls{mpc} reaches the target of \qty{10}{\centi\meter} in \qty{25}{\second}, which is twice as fast as the \gls{mimpc}, though it uses three times more thrust than the \gls{mimpc}.

The binary-informed \gls{mpc} initially exhibits a large oscillation and then settles into a small limit cycle around the target.
The resulting mean position error at the target is less than \qty{1}{\centi\meter}, while the thrusters fire on average \qty{6}{\percent} of the time.
The \gls{mimpc} reaches the target and directly converges to its limit cycle. Since this cycle is larger, the resulting average position error is approximately twice that of the binary informed \gls{mpc}. 
However, it also fires the thrusters on average for only \qty{1.4}{\percent} of the time, which is about four times less than the binary-informed \gls{mpc}. Both controllers stabilize the orientation by effectively combining the \gls{rw} and the thrusters, achieving an average error below \qty{0.5}{\degree}. 
The \gls{mimpc} achieves an error of \qty{0.18}{\degree}, which is less than half that of the binary-informed \gls{mpc}.

\section{Discussion}
The results highlight when it is advantageous to explicitly consider binary thrusters and timing constraints within a controller.
All controllers have a similar performance in reaching the target.
The continuous \gls{mpc}, however, may become unstable and can even exceed constraints in high-thrust configuration.
Furthermore, in low-thrust configuration, only the binary informed \gls{mpc} and the \gls{mimpc} are capable of reaching the target. 
During station keeping at the target, in high-thrust regimes, all compared controllers achieve the most efficient experiments with similar performance. The continuous and binary informed \gls{mpc} are even a bit more efficient in that region. This is probably due to the higher control frequency and the fact that the \gls{mimpc} does not always solve to optimality. The difference, however, is minimal and negligible.
Since the continuous \gls{mpc} in low-thrust tuning does not reach the target, only the binary informed \gls{mpc} and \gls{mimpc} are successful in station keeping with low-thrust usage. Nevertheless, in low-thrust regimes, explicit treatment of binary inputs and thruster timings within the \gls{mimpc} is advantageous, as the results indicate that \gls{mimpc} remains the most efficient controller in this operating range.
Further, the scattering of the Pareto non-optimal points indicates that in the majority of experiments, the continuous and binary informed \gls{mpc} achieved high thrust and not always minimal error, despite the weights being the same as for the \gls{mimpc}.
Consequently, achieving low-thrust usage at the target is more challenging with the binary informed and, especially, the continuous \gls{mpc}, requiring more careful tuning compared to the \gls{mimpc}.

In summary, it can be said that the binary informed \gls{mpc} is superior to the purely continuous \gls{mpc}. 
With negligible additional computing operations, the gap in efficiency between the binary informed \gls{mpc} and the \gls{mimpc} is reduced.
The results on the real system show that the binary informed \gls{mpc} works well in reality.
It even performed better than \gls{mimpc}, although it consumed more thrust. 
Therefore, a definitive assessment of efficiency on the real system is not possible, requiring more experiments and more precise tuning.
From a practical perspective, the choice of controller depends on mission constraints.
If low computational cost or a very high accuracy is required, the binary informed \gls{mpc} remains attractive as the \gls{mimpc} provides no benefit.
On the other side, if robustness and fuel efficiency are important, for example, in long-duration or resource-constrained missions, then explicitly formulating the problem as a \gls{mip} is advantageous.
Another point regarding computational complexity is that the continuous formulation still leaves “room” for additional complexity. For example, a quadratic cost function, which is in some sense more natural, could be advantageous and should be explored in future work. Other features, such as path following, a longer prediction horizon, or obstacle avoidance, could also be incorporated more easily. The \gls{mimpc}, on the other hand, is already close to the limit of what is computationally feasible.

\section{Conclusion}
This paper compared three control strategies for binary thruster systems on ESA’s \gls{reacsa} platform: a direct \gls{mimpc}, a continuous-input MPC with \gls{ds}-modulation, and a binary informed \gls{mpc} that incorporates predicted \gls{ds}-modulator firings into the \gls{ocp}.
The continuous \gls{mpc} performs reliably only at higher thrust utilizations and entails a risk of unstable behavior or even constraint violations.
The binary-informed \gls{mpc} reduces this issue by informing the \gls{ocp} formulation about the modulator’s effect within the prediction horizon. 
With substantially lower computational effort than the \gls{mimpc}, it can approach it in efficiency, while also removing the instability risk.
However, in very low-thrust regimes, explicitly managing on/off decisions and timing constraints with \gls{mimpc} reduced thruster usage for a given steady-state accuracy, thereby maximized thrust efficiency.
Experiments on the real system were consistent with the simulations. 
With identical weights, the binary informed MPC reached the goal region faster (about \SI{25}{\second} vs.\ \SI{51}{\second} for \gls{mimpc}) but at higher thruster usage (about \SI{3.7}{\percent} vs.\ \SI{1.2}{\percent}).
In steady state, the binary informed MPC produced a smaller position error (about \SI{0.85}{\centi\meter} vs.\ \SI{1.82}{\centi\meter}), while the \gls{mimpc} yielded a smaller average orientation error (about \SI{0.18}{\degree} vs.\ \SI{0.42}{\degree}).

Overall, when computational resources allow, explicitly formulating the binary decision process (\gls{mimpc}) is advantageous in low-thrust and resource-constrained scenarios.
If solver complexity is a concern, incorporating modulator-aware predictions into a continuous \gls{ocp} is a practical alternative that recovers much of the robustness observed with \gls{mimpc} while retaining efficient \glspl{lp}.
Future work can explore the potential of the binary informed MPC, for instance, through more comprehensive task specifications (such as path following or obstacle avoidance) and enhanced horizon/cost designs. 
Additionally, the amount of information provided by the modulator to the \gls{mpc}, such as when fire breaks are enforced, could be increased.

\section*{Acknowledgment}
This work is partially funded by the projects: CoEx (grant number 01IW24008) funded
by the German Federal Ministry for Education and Research, and ActGPT (grant number 01IW25002) funded by the Federal Ministry of Research, Technology and Space (BMFTR) and is supported with funds from the federal state of Bremen for setting up the Underactuated Robotics Lab (265/004-08-02-02-30365/2024-102966/2024-740847/2024).
Further, this work has been partially supported by the German Federal Ministry of Research, Technology and Space (BMFTR) under the Robotics Institute Germany (RIG).
Special thanks go to Francesca Bocconcelli, Magnus Bøgh-Larsen, and Jules Noirant for their help with the experiments at the ORGL.
Many thanks also to the Robotics Section of the European Space Agency for the wonderful collaboration.
\section*{Declaration of Use of Artificial Intelligence}
Artificial Intelligence was used to proofread and translate parts of this paper.
\bibliography{references}
\end{document}